\documentclass{article}
\usepackage[utf8]{inputenc}
\usepackage{enumerate}
\usepackage{amsmath}
\usepackage{amsthm}
\usepackage{amssymb}
\usepackage{amsfonts}
\usepackage{tikz}
\usepackage{mathtools}
\usepackage{float}
\usepackage{natbib} 
\usepackage{hyperref} 
\usepackage{csquotes}
\usepackage{graphicx}
\usepackage{dashrule}
\usepackage{xcolor}

\usetikzlibrary{arrows}

\theoremstyle{remark}

\theoremstyle{definition}
\newtheorem{definition}{Definition}
\newtheorem{theorem}{Theorem}

\newcommand{\ind}{\!\perp\!\!\!\perp}
\newcommand{\Ex}{\mathbb{E}}

\providecommand{\keywords}[1]{\textbf{\textit{Keywords---}} #1}

\title{Fairness criteria through the lens of directed acyclic graphical models}
\date{\today}
\author{
    Benjamin R. Baer\thanks{Equal contribution}, Daniel E. Gilbert\footnotemark[1], Martin T. Wells\thanks{
Wells$'$ research was partially supported by NIH U19 AI111143, NSF DMS-1611893, and Cornell's Institute for the Social Sciences project on Algorithms, Big Data, and Inequality.
} \\
    \{brb225, deg257, mtw1\}@cornell.edu \\
    Department of Statistics and Data Science \\
    Cornell University \\
}

\begin{document}

\newpage
\maketitle

\begin{abstract}
    A substantial portion of the literature on fairness in algorithms proposes, analyzes, and operationalizes simple formulaic criteria for assessing fairness. Two of these criteria, Equalized Odds and Calibration by Group, have gained significant attention for their simplicity and intuitive appeal, but also for their incompatibility. This chapter provides a perspective on the meaning and consequences of these and other fairness criteria using graphical models which reveals Equalized Odds and related criteria to be ultimately misleading. An assessment of various graphical models suggests that fairness criteria should ultimately be case-specific and sensitive to the nature of the information the algorithm processes.
\end{abstract}


\section{Introduction}
The emergence of artificial intelligence’s algorithmic tools represents one of the most important social and technological developments of the last several decades. Machine learning based scoring systems now determine creditworthiness of consumers and insurance prices \citep{robinson2014knowing} and social media metrics \citep{Duffy2016}, algorithmic hiring platforms target job advertisements and screen r\'esum\'es to calculate who should and should not be seen by human resource managers \citep{ajunwa2016hiring, gillum2019facebook}, and predictive analytics are deployed as sentencing tools in the criminal justice system \citep{eubanks2018automating}. Big data's algorithmic tools have come to play a decisive role in many aspects of our lives. However, there is concern that these algorithmic tools may lack fairness and exacerbate existing social inequalities \citep{barocas2016big, Ziewitz2016, o2017weapons, lum2017limitations}.

One might imagine that because algorithms are inherently procedural, ensuring fairness should be a simple matter of not explicitly using race or gender as features \citep{grgic2016case}. This notion of fairness has been called \emph{Fairness Through Unawareness}, and it is easy to see why it is insufficient. First of all, other features will generally \emph{redundantly encode} sensitive variables \citep{barocas2016big}. We could trivially skirt around Fairness Through Unawareness by including variables which are close proxies for gender or race like hair length or name, but even less suspicious and eminently predictive features such as zip code, language usage, or GPA will allow an algorithm to partially infer an individual's sensitive characteristics and make generalizations on those bases. Furthermore, including information about an individual's sensitive characteristics can actually serve to make a predictive algorithm \textit{more} fair, especially when there are interaction effects between sensitive characteristics and other features. The area of algorithmic fairness constitutes an attempt to move beyond Fairness Through Unawareness and develop a link between the mathematical properties of algorithms and our philosophical and intuitive notions of fairness. Unfortunately, there is little consensus on the philosophical bedrock upon which algorithmic fairness should rest \citep{binns2017fairness,chouldechova2018frontiers}.

Much of the literature on fairness in algorithms has been influenced by a controversy surrounding the Northpointe COMPAS algorithm, an algorithm for predicting criminal recidivism. \citet{angwin2016machine} analyzed the output of the algorithm and determined that its predictions were unfair due to the fact that the rate of false positives and false negatives differed significantly between racial groups. In response, Northpointe published a rejoinder arguing the criteria used by \citet{angwin2016machine} to assess fairness were nonstandard, and a proper analysis reveals that the predictions made by the COMPAS algorithm are in fact calibrated by race \citep{flores2016false}. 

Beyond merely inspiring interest in the study of algorithmic fairness, this controversy may have influenced the early direction of the field. One significant branch of the field is concerned with the development, study, comparison, and implementation of simple fairness criteria, much like the balanced-odds criterion implicit in \citet{angwin2016machine} or the calibration criterion used in \citet{flores2016false}. These fairness criteria have largely been tailored to the classification setting.

Furthermore, \citet{angwin2016machine} had access to limited information in assessing the COMPAS algorithm; the authors were able to acquire the COMPAS scores for 11,575 pretrial defendants, along with information about their criminal histories, race, and whether that defendant in fact went on to reoffend \citep{larson2016we}. However, these authors did not have access to the precise features used by the COMPAS algorithm nor the specifications of the COMPAS algorithm itself. Therefore, the authors assessed the fairness of the COMPAS algorithm by examining its false positive and false negative rates across races, which can be calculated using only the COMPAS predictions, the races of the defendants, and whether they reoffended.

Other commonly considered fairness-apt data sets have a similar form; we often desire to assess whether a predictive algorithm is fair using only information about the predictions, the observed outcomes, and the race or gender of the subjects. Perhaps for this reason, much of the early work on algorithmic fairness has centered around so-called \emph{oblivious} fairness criteria, which assess algorithms only on the basis of their outputs compared to the ground truth. The three central oblivious criteria are most often called \emph{Demographic Parity}, \emph{Equalized Odds}, and \emph{Calibration by Group}, although it is common to encounter these and related concepts under a host of names.

Two prominent strains of criticism have emerged which cast doubt on the utility of simple one-size-fits-all metrics for the fairness of algorithms. The first criticism concerns obliviousness; even alongside the introduction of Equalized Odds, \citet{hardt2016equality} note that intuitively fair and intuitively unfair algorithmic procedures can appear identical if we only compare the algorithm's output to the ground truth. Indeed, many of our intuitive notions of fairness have to do with the nature of the information used to make a prediction, rather than the outcome.

A second strain of criticism concerns incompatibility between the three oblivious fairness criteria, and thus their lack of universality. Most notably, \citet{chouldechova2017fair} and \citet{kleinberg2016inherent} proved that Calibration and Equalized Odds could not simultaneously be achieved; this recast the disagreement between ProPublica and Northpointe as a philosophical rather than statistical debate. 

Various review papers have been written which tie together the outpouring of early ideas in fairness in algorithms. In this paper, we do not intend to exhaustively catalogue the world of fairness criteria: instead we will focus on a small number of basic criteria which have received significant attention, similar to \citet{yeom2018discriminative}. For a comprehensive map of fairness measures and their relationships to one another, see \citet{mitchell2018prediction}. \citet{corbett2018measure} elucidates the incompatibility of Calibration and Equalized Odds using visualizations of outcome distributions. These authors argue that the problem of \emph{infra-marginality} suggests that Equalized Odds is a poor criterion for fairness.

We agree. We are concerned that work which generalizes and operationalizes Equalized Odds may further obscure the criterion's underlying flaws. The purpose of this article is to provide an alternate source of intuition about fairness criteria using probabilistic directed acyclic graphical models. Graphical models have been used to motivate and expose fairness criteria in other works, especially those which work with explicitly causal criteria for fairness. We believe that graphical models provide an invaluable source of intuition even in non-causal scenarios, and themselves reveal the weakness of Equalized Odds.

Using Bayesian networks, we can view fairness criteria in a way which is easily generalized beyond classification settings. Considering generalizations as defined in \citet{barocas2017fairness} of Demographic Parity, Equalized Odds, and Calibration helps to expose certain fundamental aspects of these criteria which the classification setting obscures. 

In Section \ref{bayes-nets}, we provide a brief overview of probabilistic directed graphical models and the associated causal theory. In Section \ref{measures}, we define the three oblivious fairness criteria and their generalizations. In Section \ref{scenarios}, we discuss two graphical scenarios and the implications of various fairness criteria therein. In Section \ref{understanding}, we review the incompatibility between Equalized Odds and Calibration and give a graphical view of the problem with Equalized Odds. This motivates a modified class of criteria which we call \emph{Separation by Signal}. Finally, in Section  \ref{causal}, we discuss the relationship between causality and fairness.

\section{Graphical Models}
\label{bayes-nets}

A directed acyclic graph (DAG) $\mathcal{G}$ is a pair $\{\mathbf{V},\mathbf{E}\}$ where $\mathbf{V} = \{V_1,\dots,V_n\}$ is a set of nodes and $\mathbf{E}$ is a set of directed edges, each pointing from one node to another. The \emph{acyclic} property of DAGs requires that the edges in $\mathbf{E}$ never form a \emph{directed path} leading from one variable back to itself. Let the \emph{parents} of $V_i$, $\text{Pa}(V_i)$, refer to the set of nodes which share an edge with $V_i$ such that the edge is pointing to $V_i$. 

\subsection{Probabilistic Directed Acyclic Graphical Models}
\label{bayes-nets:pdags}

A probabilistic directed acyclic graphical (PDAG) model, sometimes known as a Bayesian Network, is a pair $\{\mathcal{G},\mathcal{P}\}$ where $\mathcal{G}$ is a DAG and $\mathcal{P}$ is a probability distribution over the nodes of $\mathcal{G}$ \citep{pearl2009causality,spirtes2000causation}. Each node $V_1,\dots,V_n$ in $\mathcal{G}$ represents a random variable, and these random variables are jointly governed by the probability distribution $\mathcal{P}$. In this paper, we will consider only PDAG models which satisfy the Markov Condition. A PDAG model $\{\mathcal{G},\mathcal{P}\}$ satisfies the \emph{Markov Condition} only if the probability distribution $\mathcal{P}$ can be factored into the conditional distributions of each node given its parents. That is,
\begin{equation}
    \mathcal{P}(V_1,\dots,V_n) = \prod_{i=1}^n \mathcal{P}(V_i \mid \text{Pa}(V_i)).
\end{equation}
Node $V_1$ is considered an \emph{ancestor} of node $V_2$ if there is a directed path leading from $V_1$ to $V_2$. In that case, node $V_2$ is a \emph{descendent} of $V_1$. A node is a \emph{root} if it has no ancestors and a \emph{leaf} if it has no descendants. The random variables associated with root nodes we call \emph{exogenous} because their distribution does not depend on any of the other modelled variables.

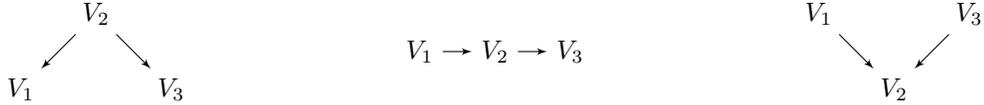
\begin{figure}
\begin{tabular}{c c c}
\centering
\begin{minipage}[c]{.333\textwidth}
\centering
     \begin{tikzpicture}
        \tikzset{edge/.style = {->,> = latex'}}
        \node (V1) at  (0,0) {$V_1$};
        \node (V2) at  (1,1) {$V_2$};
        \node (V3) at (2,0) {$V_3$};
        \draw[edge] (V2) to (V1);
        \draw[edge] (V2) to (V3);
    \end{tikzpicture}
\end{minipage} & 
\begin{minipage}[c]{.333\textwidth}
\centering
     \begin{tikzpicture}
        \tikzset{edge/.style = {->,> = latex'}}
        \node (V1) at  (0,0) {$V_1$};
        \node (V2) at  (1,0) {$V_2$};
        \node (V3) at (2,0) {$V_3$};
        \draw[edge] (V1) to (V2);
        \draw[edge] (V2) to (V3);
    \end{tikzpicture}
\end{minipage} &
\begin{minipage}[c]{.333\textwidth}
\centering
     \begin{tikzpicture}
        \tikzset{edge/.style = {->,> = latex'}}
        \node (V1) at  (0,0) {$V_1$};
        \node (V2) at  (1,-1) {$V_2$};
        \node (V3) at (2,0) {$V_3$};
        \draw[edge] (V1) to (V2);
        \draw[edge] (V3) to (V2);
    \end{tikzpicture}
\end{minipage}
\end{tabular}
\caption{In the leftmost graph, $V_2$ is a confounder, in the center graph, $V_2$ is a mediator, and in the rightmost graph, $V_2$ is a collider.}
\label{fig:nodetypes}
\end{figure}

In discussing PDAG models, three common relationships between nodes are of particular interest. Nodes $V_1$ and $V_3$ are \emph{confounded} by node $V_2$ if $V_2$ is an ancestor of both $V_3$ and $V_1$. In this case, we say there is a \emph{backdoor path} between $V_1$ and $V_3$. If $V_1$ is an ancestor of $V_2$ and $V_2$ is an ancestor of $V_3$, then $V_2$ is a \emph{mediator} of the relationship between $V_1$ and $V_3$. Finally, if $V_1$ and $V_2$ are both ancestors of $V_3$, then $V_3$ is said to be a \emph{collider} for $V_1$ and $V_2$. See Figure~\ref{fig:nodetypes} for a depiction of these relationships. 

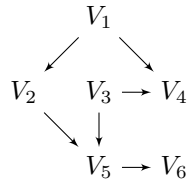
\begin{figure}
    \centering
    \begin{tikzpicture}
        \tikzset{edge/.style = {->,> = latex'}}
        \node (V1) at  (1,2) {$V_1$};
        \node (V2) at  (0,1) {$V_2$};
        \node (V3) at (1,1) {$V_3$};
        \node (V4) at (2,1) {$V_4$};
        \node (V5) at (1,0) {$V_5$};
        \node (V6) at (2,0) {$V_6$};
        \draw[edge] (V1) to (V2);
        \draw[edge] (V1) to (V4);
        \draw[edge] (V2) to (V5);
        \draw[edge] (V3) to (V4);
        \draw[edge] (V3) to (V5);
        \draw[edge] (V5) to (V6);
    \end{tikzpicture}
    \caption{In this PDAG model, nodes $V_2$ and $V_3$ are d-separated \textit{a priori}, that is, conditional on the empty set $\mathcal{S} = \varnothing$. However, conditional on the collider $\mathcal{S} = \{V_5\}$, $V_2$ and $V_3$ are d-connected. $V_2$ and $V_4$ are d-separated given any of the following sets: $\{V_1\}$,$\{V_1,V_5,V_3\}$ or $\{V_1,V_6,V_3\}.$}
    \label{fig:d-sep}
\end{figure}

We can determine certain marginal and conditional independence relationships between the random variables $V_1,\dots,V_n$ using the structure of the DAG. The nodes $V_i$ and $V_j$ are \emph{d-separated} if the structure of the DAG implies that $V_i$ and $V_j$ are (marginally) independent, i.e., $V_i \ind V_j$. A set $\mathcal{S}$ of nodes \emph{d-separates} $V_i$ and $V_j$ if the structure of the DAG implies that $V_i$ and $V_j$ are independent given the variables in $\mathcal{S}$, i.e. $V_i \ind V_j \mid \mathcal{S}$. Specifically, under the Markov Condition, the nodes $V_i$ and $V_j$ are d-separated given $S$ if $S$ blocks all paths between $V_i$ and $V_j$. A connected sequence of edges between two nodes is considered a \emph{path} regardless of the edges' orientation. A path is \emph{blocked} if either:
\begin{itemize}
    \item it contains a mediator or confounder $V_k$ where $V_k \in \mathcal{S}$, or
    \item it contains a collider $V_k$ where $V_k \not\in \mathcal{S}$ and if $V_l$ is a descendent of $V_k$, $V_l \not\in \mathcal{S}$.
\end{itemize}
Thus, conditioning on colliders (or their descendents) actually \textit{unblocks} paths and can induce dependency between marginally independent variables. See Figure~\ref{fig:d-sep} for examples. Note that while d-separation implies conditional independence, \emph{d-connection} or lack of d-separation does not necessarily imply dependence. Therefore, it is sometimes useful to make the assumption that a PDAG model is \emph{faithful}, which means that every conditional d-connection relationship in the graph implies dependence between those variables. 

\begin{figure}
    \centering
    \begin{tikzpicture}
        \tikzset{edge/.style = {->,> = latex'}}
        \node (V1) at  (1,2) {$V_1$};
        \node (V2) at  (0,1) {$V_2$};
        \node (V3) at (2,1) {$V_3$};
        \node (V4) at (1,0) {$V_4$};
        \draw[edge] (V1) to   (V2);
        \draw[edge] (V1) to   (V3);
        \draw[edge] (V2) to   (V4);
        \draw[edge] (V3) to  (V4);
    \end{tikzpicture}
    \caption{Because there are multiple paths from $V_1$ to $V_4$, this PDAG model may be unfaithful if the effect of $V_1$ on $V_4$ along one path perfectly counteracts the effect along the other path.}
    \label{unfaithful}
\end{figure}
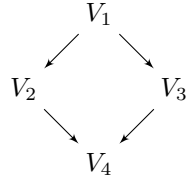

Unfaithfulness can occur because whenever there exist multiple paths leading from $V_i$ to $V_j$, the dependencies along those paths can cancel each other out. For example, consider a PDAG model associated with the DAG in Figure \ref{unfaithful}. Let $V_2 = 3V_1 + \epsilon_2$, $V_2 = 2V_3 = V_1 + \epsilon_3$, and $V_4 = -2V_2 + 3V_3 + \epsilon_4$. Then $V_4 = 3\epsilon_3 - 2\epsilon_2 + \epsilon_4$, thus $V_4$ is independent of $V_1$, despite the fact that $V_1$ and $V_4$ are d-connected. Thus this PDAG model is unfaithful. Note, however, it is unusual for path effects to precisely cancel each other except when variables are carefully constructed to do so.

\subsection{Causality}
\label{bayes-nets:causality}

Strictly speaking, the directed edges in PDAG models need not have any causal interpretation, as long as they are consistent with the conditional dependencies in $\mathcal{P}$. However, DAGs are not fully determined by their associated probability distributions: a given probability distribution is usually consistent with multiple DAGs with differently oriented edges. Thus it is natural to use a PDAG model to convey causal meaning, so that an edge pointing from $V_i$ to $V_j$ then means that $V_i$ has a causal effect on $V_j$. When PDAG models are used in the context of causal inference, they are often called \emph{Structural Causal Models} and the associated DAG may be called a \emph{Causal Graph}.

Pearl's theory of causality addresses two types of causal questions: questions about the effects of manipulating variables and questions about counterfactual states of affairs \citep{pearl2009causality}. We will focus on inferences about variable manipulations. Pearl uses a construct called the \emph{do() operator} to express \emph{do-statements} such as $\mathcal{P}(V_1 = v_1 \mid \text{do}(V_2 = v_2))$. This statement can be interpreted as ``the probability that $V_1 = v_1$ when we intervene to set $V_2 = v_2$.'' Formally, to \emph{intervene} on the variable $V_2$ by setting it to $v_2$ means to construct an alternate PDAG model in which the edges between $V_2$ and its ancestors are deleted and the distribution of the root $V_2$ is set to be a point mass at $v_2$.  

\begin{figure}
 \centering
    \begin{tikzpicture}
        \tikzset{edge/.style = {->,> = latex'}}
        \tikzset{vertex/.style = {shape=rectangle,draw,minimum size=1.5em}}
        \node[vertex,dotted] (V1) at  (1,2) {$V_1$};
        \node (V2) at  (0,1) {$V_2$};
        \node (V3) at (2,1) {$V_3$};
        \draw[edge] (V1) to (V2);
        \draw[edge] (V1) to (V3);
        \draw[edge] (V2) to (V3);
    \end{tikzpicture}
    \caption{In this PDAG model, $V_2$ and $V_3$ are confounded by the unobserved variable $V_1$; this will render an expression such as $\mathcal{P}(V_3 \mid \text{do}(V_2))$ unidentifiable.}
    \label{unobserved}
\end{figure}
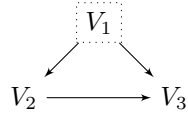

These do-statements can sometimes be resolved into equivalent \emph{see-statements} using Pearl's three rules of do-calculus, which are consequences of the Markov Condition. See-statements are expressions which may involve various conditional probabilities, but do not contain the do-operator. Thus, see-statements can be evaluated using only information about the joint probability distribution $\mathcal{P}$ of the variables in the original PDAG. Note that in some cases, we may include \emph{unobserved} variables in a PDAG model. We will indicate that a variable is unobserved in a DAG with a dotted outline as in Figure \ref{unobserved}. When do-statements cannot be resolved into see-statements depending only on observed variables, they are called \emph{unidentifiable}. 

For an accessible and more complete introduction to PDAG models and Pearl's causal theory, see \citet{pearl2016causal, pearl2009causal}. The purpose of our use of PDAG models in this paper is mostly to provide intuition regarding sets of variables with various conditional dependency relationships. However, in Section \ref{causal} we will discuss certain aspects of fairness which require a properly causal treatment. 

\section{Three Criteria for Fairness}
\label{measures}

We will review the definitions of three prominent fairness criteria: Independence, Separation, and Sufficiency, which we will examine through the lens of PDAG models in this paper. Let $Y$ be a \emph{response}, an outcome of interest measured for an individual. For example, $Y$ could be whether an individual will repay a loan, or whether she will click on an advertisement. Let $A$ be a \emph{sensitive characteristic}, a categorical variable indicating that individual's class with respect to a fairness-apt feature such as race or gender. Let $R$ be a \emph{prediction}, an estimated response for that individual.

If we select an individual at random from the population, the quantities $Y$, $A$, and $R$ can be modelled as random variables. We are concerned with assessing whether $R$ is a fair prediction. The three fairness criteria we examine in this paper are \emph{oblivious} criteria, which means they assess only the joint distribution of the tuple $(A,R,Y)$ \citep{hardt2016equality}. In other words, these criteria are not concerned with the functional form of $R$ or the information upon which it is based; they treat $R$ as a black box.

Most of the work that has been done on fairness criteria for machine learning has considered the classification setting, in which $Y$ is a categorical (and often binary) variable. Therefore it is no surprise that each of the three fairness criteria defined in this section were first introduced as criteria for assessing classifiers. These fairness criteria for the classification setting are known as Demographic Parity, Equalized Odds, and Calibration by Group. However, \citet{barocas2017fairness} offers sensible generalizations of these three criteria to settings with arbitrary, possibly continuous $R$ and $Y$. Here we will introduce both the original and generalized versions of each of the three criteria.  

\subsection{Demographic Parity and Independence}
\label{measures:indep}

As a starting point for assessing the fairness of a prediction algorithm, we may ask whether the algorithm is making systematically different predictions for different groups. Suppose $A$ is a categorical sensitive characteristic taking values in the set $\mathbb{A}$. Considering only the binary classification case for the moment, suppose that the response $Y \in \{0,1\}$ and the prediction $R \in \{0,1\}$. 
\begin{definition}
\label{DP}
    A prediction $R$ satisfies Demographic Parity if $$P(R = 1 \mid A = a) = P(R = 1 \mid A = a')$$ for every sensitive characteristic $a,a' \in \mathbb{A}$.
\end{definition}
In the binary case, this is equivalent to the statement $R \ind A$. Therefore, the natural generalization of Demographic Parity as suggested by \citet{barocas2017fairness} is as follows. For arbitrary random variables $R$, $A$ and $Y$,
\begin{definition}
\label{Ind}
    A prediction $R$ satisfies Independence if $R \ind A$.
\end{definition}
\noindent This is a strong criterion in the sense that it requires that no aspect of the distribution of $R$ depend on $A$. A weaker form of Independence could require only that the expected value and possibly the variance of $R$ not depend on $A$. See \citet{johndrow2019algorithm,calders2009building,calders2010three,del2018obtaining,hacker2017continuous} for methods for achieving full or partial Independence in predictions.

\subsection{Equalized Odds and Separation}
\label{measures:sep}

The Independence criterion does not take the response $Y$ into account; that is, it enforces equality of the distributions of the prediction $R$ across the protected characteristic $A$ even when the distribution of the response $Y$ may differ across protected classes. For binary classifiers, \citet{hardt2016equality} argues:
\begin{displayquote}
    Demographic Parity is seriously flawed on two counts. First, it doesn't ensure fairness. The notion permits that we accept the qualified applicants in one demographic, but random individuals in another, so long as the percentages of acceptance match. This behavior can arise naturally, when there is little or no training data available for one of the demographics. Second, demographic parity often cripples the utility that we might hope to achieve, especially in the common scenario in which an outcome to be predicated, e.g. whether the loan will be defaulted, is correlated with the protected attribute. 
\end{displayquote}
In light of this, \citet{hardt2016equality} proposes an alternative criterion for fairness. Suppose $Y \in \{0,1\}$ and the prediction $R \in \{0,1\}$. 
\begin{definition}
\label{EO}
    A prediction $R$ satisfies Equalized Odds if $$P(R = 1 \mid Y = y, A = a) = P(R = 1 \mid Y = y, A = a')$$ for every sensitive characteristic $a,a' \in \mathbb{A}$ and response $y \in \{0,1\}$.
\end{definition}

\citet{hardt2016equality} argues that when $Y$ and $A$ are not independent, $Y$ itself does not satisfy Demographic Parity, and therefore nor would a ``perfect'' classifier $R=Y$. On the other hand, a ``perfect'' classifier $R=Y$ will always satisfy Equalized Odds. Thus unless we are attempting to modify our predictions as a form of affirmative action, Equalized Odds seems to have an advantage over Demographic Parity. In Section \ref{understanding:schmeparation} we argue that this intuition regarding ``perfect'' classifiers is an artifact of the discrete classification setting and its motivation has less force in arbitrary regression settings. Thus we will consider a general form of Equalized Odds offered by \citet{barocas2017fairness}.
\begin{definition}
\label{Separation}
    A prediction $R$ satisfies Separation if $R \ind A \mid Y$.
\end{definition}
\noindent Like Independence, this is a strong criterion, and can be relaxed by requiring only that the conditional expectation $\Ex(R \mid Y)$ and possibly the conditional variance $\text{Var}(R \mid Y)$ do not depend on $A$. See \citet{hardt2016equality,pleiss2017fairness,donini2018empirical,zafar2017fairness,corbett2017algorithmic} for expositions of Separation-like criteria and methods for enforcing them. 

\subsection{Calibration by Group and Sufficiency}
\label{measures:suff}

Calibration itself is not a fairness concept; it is a popular criterion for assessing an aspect of the performance of predicted probabilities \citep{lichtenstein1981calibration}. Consider the case where the response $Y \in \{0,1\}$ is binary and the predicted probability that $Y=1$ is $P \in [0,1]$. 
\begin{definition}
\label{def:calibration}
    A predicted probability $P$ satisfies \emph{Calibration} if $P(Y=1 \mid P = p) = p$ for any $p \in [0,1]$.
\end{definition}
\noindent This would suggest that classifications generated by a calibrated predicated probability are trustworthy in the sense that a practitioner has no incentive to make post-hoc adjustments to compensate for known biases in ranges of the prediction.

A related criterion, which is directly relevant to fairness, is whether a predicted probability is calibrated across subpopulations, that is, whether a given predicted probability has the same meaning when generated for individuals from different subpopulations. Suppose $Y \in \{0,1\}$ and $P \in [0,1]$. 
\begin{definition}
\label{def:calibration.by.group}
    A predicted probability $P$ satisfies \emph{Calibration by Group} if $P(Y = 1 \mid P = p, A = a) = p$ for each sensitive attribute $a \in \mathbb{A}$ and probability $p \in [0,1]$.
\end{definition}
Calibration by Group is intuitively appealing because if it is not satisfied, some individuals' predictions must deviate from the model-grounded truth in a manner depending on their group membership. That is, a predicted probability $P$ which satisfies Calibration by Group has equal performance across the sensitive attribute. Indeed, Calibration by Group is a combination of Calibration and lack of dependence on the sensitive attribute $A$. The lack of dependence can be isolated, in terms of a prediction $R \in \{0, 1\}$, through the following definition.
\begin{definition}
\label{def:predictive.parity}
    A prediction rule $R \in \{0, 1\}$ satisfies \emph{Predictive Parity} if $P[Y = 1 \mid R = r, A = a] = P[Y=1 \mid R=r]$ where the prediction $r \in \{0,1\}$ and the sensitive characteristic $a \in \mathbb{A}$.
\end{definition}
Predictive Parity was discussed and coined by \citet{chouldechova2017fair}. \citet{barocas2017fairness} present a natural generalization of predictive parity for a not necessarily binary $Y$ and $R$.
\begin{definition}
\label{def:sufficiency}
    A prediction $R$ satisfies Sufficiency if $Y \ind A \mid R$.
\end{definition}

\section{Fairness Criteria in Two Scenarios}
\label{scenarios}

To supplement the basic motivations of these three fairness criteria, we will discuss their implications in two prediction scenarios. We will find that these various criteria do not represent equally viable choices with different subjective implications, but rather that certain criteria are operational in certain scenarios, and seemingly meaningless in others. Independence has clear use cases and represents an assumption about the relationships between the sensitive characteristic and the response, or else a desire to impose a regime of special intervention in favor of a particular group. Sufficiency, on the other hand, serves more as a measure of the extent to which a prediction takes advantage of all of the information relevant in predicting the response. And finally, Separation does have meaningful use cases in esoteric scenarios such as Scenario 2. However, in most scenarios of interest, Separation has counterproductive and destructive implications.

\subsection{Scenario 1: Loan Repayment}
\label{scenarios:loan}
\begin{figure}
\begin{tabular}{c c}
\begin{minipage}{.4\textwidth}
    \begin{center}
    \begin{tikzpicture}
        \tikzset{edge/.style = {->,> = latex'}}
        \node (A) at  (0,1) {$A$};
        \node (X1) at  (0,0) {$X_1$};
        \node (X2) at (1,1) {$X_2$};
        \node (X3) at (2,0) {$X_3$};
        \node (Y) at (2,1) {$Y$};
        \draw[edge] (A) to (X1);
        \draw[edge] (A) to (X2);
        \draw[edge] (X2) to (Y);
        \draw[edge] (X3) to (Y);
    \end{tikzpicture}
    \end{center}
\end{minipage}
&
\begin{minipage}{.4\textwidth}
    \begin{tabular}{c l}
        $A$ & Race \\
        $Y$ & Repays Loan? \\
        $X_1$ & Applicant's Hair Color \\
        $X_2$ & Applicant's Credit Rating \\
        $X_3$ & Loan Interest Rate
    \end{tabular}
\end{minipage}
\end{tabular}
\caption{The random variables in Scenario 1: various features and their relationship to race $A$ and loan repayment $Y$.}
\label{fig:loan}
\end{figure}
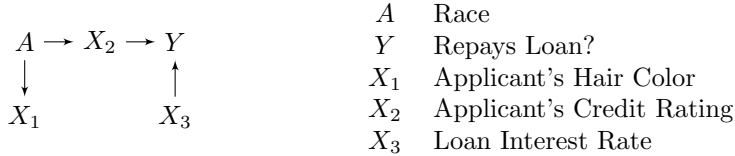

We wish to predict whether an individual will repay a loan. Suppose we model the situation using the PDAG in Figure~\ref{fig:loan}.
    
We consider three features with different conditional dependency structures encoded by this PDAG model. The applicant's hair color $X_1$ is a descendant of her race $A$ and is not a mediator of the effect of $A$ on the whether she repays the loan, $Y$. Perhaps for this reason, it seems intuitively unfair (and is illegal in some places \citep{stowe2019new}) to determine an applicant's loan premium based on her hair color. However, in some cases it may be tempting to do so because there is a backdoor path connecting $X_1$ and $Y$, thus $X_1$ and $Y$ are statistically marginally dependent. The nature of $X_1$ illustrates one shortcoming of \emph{Fairness Through Unawareness}, which demands we do not use $A$ itself as an input into our prediction. It would be no violation of Fairness Through Unawareness to use $X_1$ alone to predict $Y$. However, if we do so, we are merely taking advantage of the backdoor path through $A$; in other words, we are using a noisy version of $A$ to predict $Y$ rather than $A$ itself. 

The applicant's credit rating $X_2$ is a mediator between $A$ and $Y$. Therefore, it is statistically dependent on race, but is also predictive of whether the applicant will repay the loan even after taking race into consideration. While an applicant's credit rating is a natural feature for predicting loan repayment, it redundantly encodes race to some extent. Finally, the interest rate of the loan $X_3$ influences $Y$ but is marginally independent of race. Therefore, $X_3$ itself seems to be an innocuous prediction. 

We now assess the implications of the Independence, Separation, and Sufficiency criteria in this context.

\subsubsection{Independence}
\label{scenarios:loan:ind}

The Independence criterion requires that our prediction $R$ is marginally independent of the applicant's race $A$. Thus, assuming that our PDAG model is faithful, we can achieve Independence only by positioning $R$ such that it is unconditionally d-separated from $A$. The applicant's hair color $X_3$ is itself d-separated from $A$, therefore any prediction $R$ which is a descendant only of $X_3$ will always satisfy Independence. 

\begin{figure}
    \centering
    \begin{tikzpicture}
        \tikzset{edge/.style = {->,> = latex'}}
        \node (A) at  (0,1) {$A$};
        \node (X1) at  (0,0) {$X_1$};
        \node (X2) at (1,1) {$X_2$};
        \node (X3) at (2,0) {$X_3$};
        \node (Y) at (2,1) {$Y$};
        \node (R) at (3,0) {$R$};
        \draw[edge] (A) to (X1);
        \draw[edge] (A) to (X2);
        \draw[edge] (X2) to (Y);
        \draw[edge] (X3) to (Y);
        \draw[edge] (X3) to (R);
    \end{tikzpicture}
    \caption{This prediction $R$ is d-separated from $A$ and therefore satisfies Independence.}
    \label{fig:loan:ind}
\end{figure}
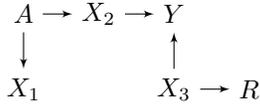

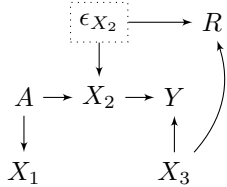
\begin{figure}
    \centering
    \begin{tikzpicture}
        \tikzset{edge/.style = {->,> = latex'}}
        \tikzset{vertex/.style = {shape=rectangle,draw,minimum size=1.5em}}
        \node (A) at  (0,1) {$A$};
        \node (X1) at  (0,0) {$X_1$};
        \node (X2) at (1,1) {$X_2$};
        \node (X3) at (2,0) {$X_3$};
        \node (Y) at (2,1) {$Y$};
        \node[vertex,dotted] (EX2) at (1,2) {$\epsilon_{X_2}$};
        \node (R) at (2.5,2) {$R$};
        \draw[edge] (A) to (X1);
        \draw[edge] (A) to (X2);
        \draw[edge] (X2) to (Y);
        \draw[edge] (X3) to (Y);
        \draw[edge,bend right] (X3) to (R);
        \draw[edge] (EX2) to (R);
        \draw[edge] (EX2) to (X2);
    \end{tikzpicture}
    \caption{The prediction $R$ depends on only the part of $X_2$ independent of race $A$.}
    \label{fig:loan:ind2}
\end{figure}

However, a prediction $R$ which descends from either the applicant's hair color $X_1$ or credit rating $X_2$ would fail to satisfy Independence. Nonetheless, there may remain valuable information within $X_2$ which may help us predict $Y$ while maintaining Independence. We can extract this information by constructing a model which can be represented by the PDAG in Figure~\ref{fig:loan:ind2}.

By decomposing $X_2$ into a component which depends on $A$ and an exogenous component $\epsilon_{X_2}$ which is marginally independent of $A$, we can construct a prediction $R$ which uses more information but is still d-separated from $A$ and thus satisfies Independence. While $X_2$ is observed in our model, this exogenous component $\epsilon_{X_2}$ is unobserved, and must be recovered in a model-specific manner. Consider a simple case, in which $X_2 \mid A \sim \mathcal{N}(\mu_A,1)$. Then $\epsilon_{X_2} = (X_2 - \mu_A) \sim \mathcal{N}(0,1)$ would satisfy the conditional independencies encoded by this PDAG model, and thus we may safely allow our prediction $R$ to depend on $\epsilon_{X_2}$.

In the context of this scenario, satisfying Independence while using information about $X_2$ entails that we use a version of $X_2$ which is de-meaned by race. That is, we would use as a feature an individual's credit rating relative only to others of the same race. 

This procedure may seem justifiable if credit ratings are themselves racially biased and thus an inaccurate indicator of an individual's likelihood of repaying the loan. However this is untrue by assumption in our model, because $Y$ is conditionally independent of $A$ given $X_2$. Thus in this case, the procedure of de-meaning credit ratings by race to satisfy Independence should be interpreted as a special modification of the prediction $R$ to benefit a particular (likely disadvantaged) group. If credit ratings are in fact racially biased, we may use a modification of the model in Scenario 1.

\begin{figure}
    \centering
    \begin{tikzpicture}
        \tikzset{edge/.style = {->,> = latex'}}
        \tikzset{vertex/.style = {shape=rectangle,draw,minimum size=1.5em}}
        \node (A) at  (0,1) {$A$};
        \node (X1) at  (0,0) {$X_1$};
        \node (X2) at (1,1) {$X_2$};
        \node (X3) at (2,0) {$X_3$};
        \node (Y) at (2,1) {$Y$};
        \node[vertex,dotted] (EX2) at (1,2) {$\epsilon_{X_2}$};
        \node (R) at (2.5,2) {$R$};
        \draw[edge] (A) to (X1);
        \draw[edge] (A) to (X2);
        \draw[edge] (X3) to (Y);
        \draw[edge,bend right] (X3) to (R);
        \draw[edge] (EX2) to (R);
        \draw[edge] (EX2) to (X2);
        \draw[edge] (EX2) to (Y);
    \end{tikzpicture}
    \caption{A modification of Scenario 1 in which the only part of $X_2$ which contributes to the response $Y$ is the noise $\epsilon_{X_2}$, independent of $A$.}
    \label{fig:loan:ind3}
\end{figure}
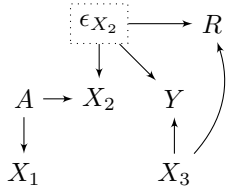

In this modified scenario, $X_2$ is a racially biased credit rating which unfairly modifies information about an applicant's true driver of default risk, $\epsilon_{X_2}$. Under these assumptions, we can achieve Independence without sacrificing any predictive accuracy. 

Finally, in Scenario 1, no information about an applicant's hair color $X_1$ can be productively used while maintaining Independence. Any exogenous components which resulted from a decomposition of $X_1$ would be independent of $Y$; in this scenario, Independence bars us from considering an applicant's hair color entirely. 

\subsubsection{Separation}
\label{scenarios:loan:sep}

The Separation criterion requires that our prediction $R$ is independent of an applicant's race $A$ conditional on whether she does in fact repay the loan, $Y$. We have seen that Independence is a strong criterion that requires that $R$ has no dependency on $A$ despite the fact that $Y$ itself is dependent on $A$. In Scenario 1, we may desire a criterion which allows us to take into account more information about the applicant's credit rating $X_2$ (because $X_2$ has a direct and perhaps causal relationship with $Y$). Nonetheless, we may we still desire that our criterion prohibits the use of spurious information like hair color, $X_1$. 

However, in this scenario, Separation does no such thing. To see this clearly, we will consider all prediction rules which are descendants of only one feature. Let $R_j$ denote an arbitrary prediction rule which depends on only the feature $X_j$, for each $j = 1, 2, 3$.

\begin{figure}
\begin{tabular}{c c c}
\centering
\begin{minipage}[c]{.333\textwidth}
\centering
     \begin{tikzpicture}
        \tikzset{edge/.style = {->,> = latex'}}
        \node (A) at  (0,1) {$A$};
        \node (X1) at  (0,0) {$X_1$};
        \node (X2) at (1,1) {$X_2$};
        \node (X3) at (2,0) {$X_3$};
        \node (Y) at (2,1) {$Y$};
        \node (R) at (-1,0) {$R_1$};
        \draw[edge] (A) to (X1);
        \draw[edge] (A) to (X2);
        \draw[edge] (X2) to (Y);
        \draw[edge] (X3) to (Y);
        \draw[edge] (X1) to (R);
    \end{tikzpicture}
\end{minipage} & 
\begin{minipage}[c]{.333\textwidth}
\centering
     \begin{tikzpicture}
        \tikzset{edge/.style = {->,> = latex'}}
        \node (A) at  (0,1) {$A$};
        \node (X1) at  (0,0) {$X_1$};
        \node (X2) at (1,1) {$X_2$};
        \node (X3) at (2,0) {$X_3$};
        \node (Y) at (2,1) {$Y$};
        \node (R) at (1,0) {$R_2$};
        \draw[edge] (A) to (X1);
        \draw[edge] (A) to (X2);
        \draw[edge] (X2) to (Y);
        \draw[edge] (X3) to (Y);
        \draw[edge] (X2) to (R);
    \end{tikzpicture}
\end{minipage} &
\begin{minipage}[c]{.333\textwidth}
\centering
     \begin{tikzpicture}
        \tikzset{edge/.style = {->,> = latex'}}
        \node (A) at  (0,1) {$A$};
        \node (X1) at  (0,0) {$X_1$};
        \node (X2) at (1,1) {$X_2$};
        \node (X3) at (2,0) {$X_3$};
        \node (Y) at (2,1) {$Y$};
        \node (R) at (3,0) {$R_3$};
        \draw[edge] (A) to (X1);
        \draw[edge] (A) to (X2);
        \draw[edge] (X2) to (Y);
        \draw[edge] (X3) to (Y);
        \draw[edge] (X3) to (R);
    \end{tikzpicture}
\end{minipage}
\end{tabular}
\caption{Three examples of predictions, each where the prediction rule $R_j$ depends only on the feature $X_j$.}
\label{fig:loan:sep1}
\end{figure}
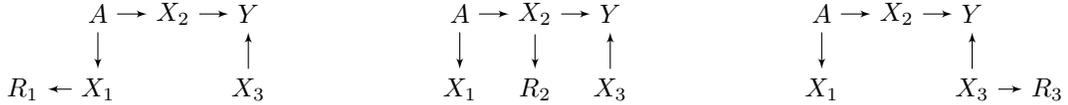

Any prediction $R_1$ depending on only $X_1$ violates Separation as expected, since $Y$ does not d-separate $R_1$ and $A$. But this is also the case for $R_2$: $Y$ does not block the path between $R_2$ and $A$. Even more surprisingly, because $Y$ is a collider, conditioning on $Y$ actually \textit{unblocks} the path between $R_3$ and $A$. Therefore even though the interest rate of the loan $X_3$ was eligible for use under Independence due to its independence with $A$, it cannot be used under Separation. Extracting components from the features $X_1$, $X_2$ or $X_3$ is also futile; any graphical descendent or ancestor of these features would still be d-connected to $A$ given $Y$.  

\begin{figure}
\centering
\begin{tikzpicture}
    \tikzset{edge/.style = {->,> = latex'}}
    \node (A) at  (0,1) {$A$};
    \node (X1) at  (0,0) {$X_1$};
    \node (X2) at (1,1) {$X_2$};
    \node (X3) at (2,0) {$X_3$};
    \node (Y) at (2,1) {$Y$};
    \node (R) at (1,0) {$R$};
    \node (Z) at (1,-1) {$Z$};
    \draw[edge] (A) to (X1);
    \draw[edge] (A) to (X2);
    \draw[edge] (A) to (R);
    \draw[edge] (X2) to (Y);
    \draw[edge] (X3) to (Y);
    \draw[edge] (X2) to (R);
    \draw[edge] (Z) to (R);
\end{tikzpicture}
\caption{The prediction $R$ takes as input both $A$ and $X_2$, whose effects conspire to violate faithfulness and make $R \ind A \mid Y$.}
\label{fig:loan:sep2}
\end{figure}
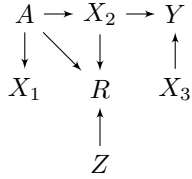

In general, Separation prevents us from constructing any predictions which are descendants of $X_1$, $X_2$ or $X_3$ in a faithful PDAG model. We can, however, induce a violation of faithfulness to force independence between $R$ and $A$. To do so, we must let $R$ be a direct descendent of $A$. For example, considering for simplicity a prediction using information about credit rating, $X_2$, suppose we the PDAG model contains the following Gaussian linear model:
\begin{align*}
    Y \mid X_2 &\sim \mathcal{N}(\beta X_2, \sigma^2), \\
    X_2 \mid A &\sim \mathcal{N}(\mu_A,\sigma_A^2).
\end{align*}
As throughout this paper, assume that all of the model parameters, $\beta$, $\sigma^2$, $\mu_A$, and $\sigma_A^2$ are known. Then, by basic properties of the multivariate normal distribution \citep{moser1996linear}, 
\begin{align*}
    X_2 \mid Y, A &\sim \mathcal{N}\left((1-\rho_A)\mu_A + \rho_A \frac{Y}{\beta}, \sigma^2 \rho_A \right),
\end{align*}
where $\rho_A = \frac{\beta^2\sigma_A^2}{\beta^2\sigma_A^2 + \sigma^2}$. 
As we may have anticipated due to the structure of the PDAG model, the mean and variance of $X_2$ given $Y$ still depend on race $A$. However, we can use this conditional distribution to construct the optimal linear prediction $R$ which cancels out these dependencies by explicitly taking into account the race $A$. It is:
\begin{align*}
    R(X_2) = \beta \left(\frac{X_2 - (1-\rho_A)\mu_A}{\rho_A} \right) + Z,
\end{align*}
where $Z \sim \mathcal{N}(0,(c - \frac{1}{\rho_A})\sigma^2)$ is an independent source of noise, and $c = \max\{\frac{1}{\rho_a}\}$ across all races, $a \in \mathbb{A}$. Thus, $R$'s dependencies are encoded by the new PDAG model which is not faithful, depicted in Figure~\ref{fig:loan:sep2}.

\begin{figure}
    \centering
    \includegraphics[scale=.4]{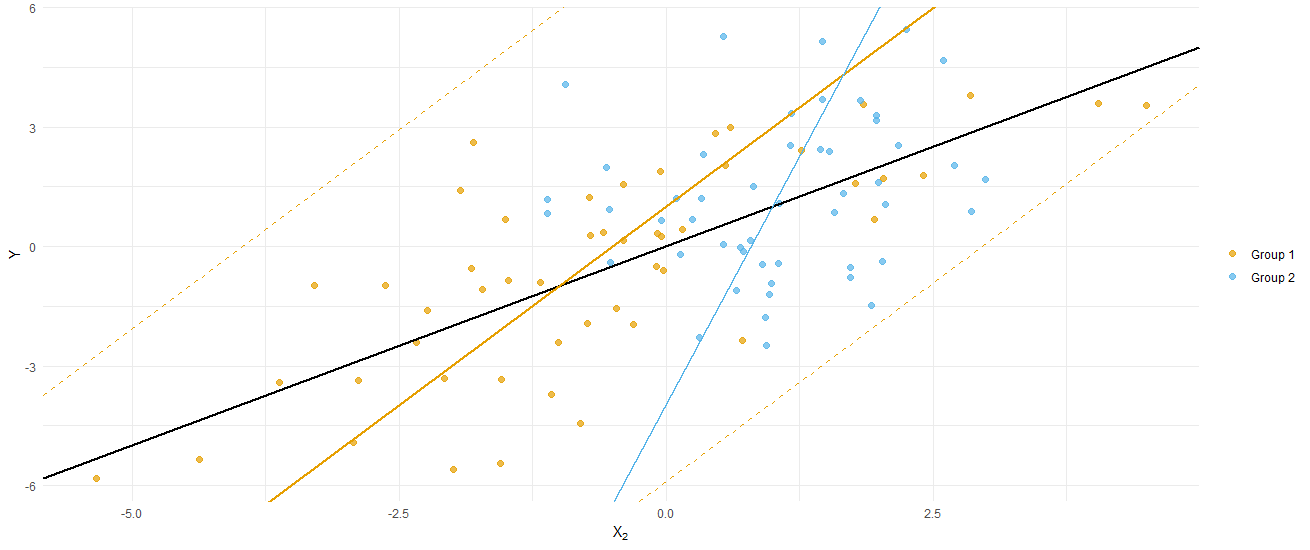}
     \\
    \caption{Here, the mean and variance of the distribution of $X_2$ differs between groups. In Group 1 (\textcolor{orange}{$\bullet$}), $X_2$ has a lower mean and greater variance than Group 2 (\textcolor{cyan}{$\bullet$}). However, for both groups, $Y = \beta X_2 + \epsilon$. That is, the known regression line (\rule[.5ex]{1cm}{1pt}) accurately captures the signal component of $Y$ for both groups. Nonetheless, the modified predictions satisfying Separation for Group 1 (\textcolor{orange}{\rule[.5ex]{1cm}{1pt}}) and Group 2 (\textcolor{cyan}{\rule[.5ex]{1cm}{1pt}}) differ and greatly deviate from the true regression line. Furthermore, to achieve equal conditional prediction variance between groups, we must randomize the predictions for Group 1; the lighter dotted lines (\textcolor{orange}{\hdashrule[.5ex]{1cm}{.25pt}{.75mm}}) indicate two-standard-deviation bounds for the randomized predictions.}
    \label{fig:sep.visualization}
\end{figure}

Note that $R(X_2) \mid Y \sim \mathcal{N}(Y,c\sigma^2)$, and this distribution does not depend on $A$. However, this prediction rule $R$ is suspicious perhaps most notably because it requires that the addition of Harrison Bergeron-esque noise to to the predictions for certain individuals in order to achieve parity in prediction error across groups. The inclusion of additional noise is similar to a result found in \citet{pleiss2017fairness} for discrete classifiers. A depiction of the prediction $R$ above is in Figure~\ref{fig:sep.visualization}.

Thus, in Scenario 1, Separation does not seem to be a natural criterion for fairness because it suggests only counterproductive procedures for constructing estimators. However, we will show that in Scenario 2, Separation sometimes has the power to discriminate between subjectively different prediction rules.

\subsubsection{Sufficiency}
\label{scenarios:loan:suff}

The Sufficiency criterion demands that whether an applicant will repay her loan $Y$ is independent of her race $A$ conditional on the prediction $R$. Because $R$ will always be a graphical descendent of some subset of the features, $X_1$, $X_2$, and $X_3$, we can see from Figure \ref{fig:loan:sep1} that we cannot achieve Sufficiency merely by carefully choosing $R$'s arguments. Conditioning on $R$ will never d-separate $Y$ from $A$.

However, we can see that the applicant's credit rating $X_2$ d-separates $Y$ from $A$, and therefore any invertible function $R(X_2)$ will satisfy Sufficiency. 
This argument will not generally work when the prediction rule $R$ is a function of more than one feature, though, since it is not generally possible to invert such a function. Indeed, in order to construct a prediction rule $R(X_2, X_3)$ which satisfies Separation, the prediction $R$ must explicitly block the path between $A$ and $Y$. This can be done when the response $Y$ only depends on the features $X_2$ and $X_3$ through a parameter $\theta(X_2, X_3)$: in this case, the predictor $R = \theta$ satisfies Sufficiency. Of course, the parameter $\theta$ which controls the way in which the probability of loan repayment depends on the loan interest rate $X_3$ and an applicant's credit rating $X_2$ is not known in practice, so this exact predictor is not available for use. This argument further shows that, in this case, the only way that prediction rules which are estimated from training data can be Sufficient is through their recovery of the signal.

\subsection{Scenario 2: Job Advertisement}
\label{scenarios:job}
\begin{figure}
\begin{tabular}{c c}
\begin{minipage}{.4\textwidth}
    \begin{center}
    \begin{tikzpicture}
        \tikzset{edge/.style = {->,> = latex'}}
        \node (A) at  (0,1) {$A$};
        \node (X1) at  (0,0) {$X_1$};
        \node (Y) at (1,1) {$Y$};
        \node (X2) at (2,1) {$X_2$};
        
        \draw[edge] (A) to (Y);
        \draw[edge] (A) to (X1);
        \draw[edge] (Y) to (X2);
    \end{tikzpicture}
    \end{center}
\end{minipage}
&
\begin{minipage}{.4\textwidth}
    \begin{tabular}{c l}
        $A$ & Gender \\
        $Y$ & Is Programmer? \\
        $X_1$ & Visited \texttt{pinterest.com}? \\
        $X_2$ & Visited \texttt{stackexchange.com}? \\
    \end{tabular}
\end{minipage}
\end{tabular}
\caption{The random variables in Scenario 2: various features and their relationship to gender $A$ and progamming employment $Y$.}
\label{fig:job}
\end{figure}
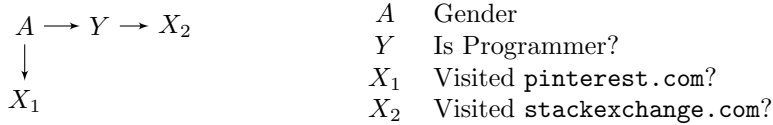

In this scenario, modelled after a similar scenario from \citet{barocas2017fairness}, we are looking to serve advertisements for a programming job to web users who are likely to be programmers. To predict whether or not the user is a programmer, we use information about his or her browsing history. For a real life example of issues in fairness which may arise from serving job advertisements, see \citet{gillum2019facebook}. Suppose we model the relationship between measurements on an individual user using the PDAG in Figure~\ref{fig:job}.

We assume $X_1$, a variable indicating whether a user has visited \texttt{pinterest.com}, has no relationship to whether or not the individual is a programmer except by virtue of the information it encodes about gender. The novelty of this scenario is that we observe $X_2$, a variable indicating whether a user has visited \texttt{stackexchange.com}, which we assume is a graphical descendant of whether a user is a programmer, $Y$. According to this model, a user's gender affects the likelihood that he or she is a programmer, but a programmer has a certain likelihood of visiting \texttt{stackexchange.com} regardless of his or her gender.

Therefore if we do not wish to target users for this employment advertisement based on gender, $X_1$ is a suspicious candidate, but $X_2$ may reasonably be considered fair game. We now examine the implications of the oblivious criteria in this scenario.

\subsubsection{Independence}
\label{scenarios:job:ind}

As in Scenario 1, if the prediction $R$ depends on whether a user has visited \texttt{pinterest.com}, $X_1$, $R$ will violate Independence, and any information in $X_1$ which is independent of a user's gender $A$ will also be independent of whether he or she is a programmer, $Y$. 
Furthermore, because $X_2$ is a descendant of $Y$ with no backdoor connection to $A$, any component of whether the user has visited \texttt{stackexchange.com}, $X_2$, which is independent of $A$ will also be independent of $Y$. Therefore, we cannot construct any non-trivial predictions $R$ in Scenario 2 which satisfy Independence. 

\subsubsection{Separation}
\label{scenarios:job:sep}

On the other hand, Scenario 2 is where Separation shines. Consider again estimators which depend on only one feature: let $R_1$ denote an arbitrary prediction rule which depends on only $X_1$, and, likewise, let $R_2$ denote an arbitrary prediction rule which depends on only $X_2$.

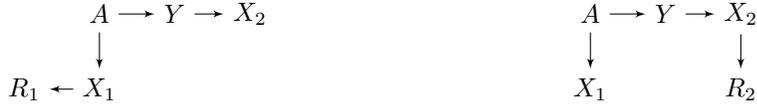
\begin{figure}
\begin{tabular}{c c c}
\centering
\begin{minipage}[c]{.45\textwidth}
\centering
     \begin{tikzpicture}
        \tikzset{edge/.style = {->,> = latex'}}
        \node (A) at  (0,1) {$A$};
        \node (X1) at  (0,0) {$X_1$};
        \node (Y) at (1,1) {$Y$};
        \node (X2) at (2,1) {$X_2$};
        \node (R) at (-1,0) {$R_1$};
        \draw[edge] (A) to (Y);
        \draw[edge] (A) to (X1);
        \draw[edge] (Y) to (X2);
        \draw[edge] (X1) to (R);
    \end{tikzpicture}
\end{minipage} & 
\begin{minipage}[c]{.45\textwidth}
\centering
     \begin{tikzpicture}
        \tikzset{edge/.style = {->,> = latex'}}
        \node (A) at  (0,1) {$A$};
        \node (X1) at  (0,0) {$X_1$};
        \node (Y) at (1,1) {$Y$};
        \node (X2) at (2,1) {$X_2$};
        \node (R) at (2,0) {$R_2$};
        \draw[edge] (A) to (Y);
        \draw[edge] (A) to (X1);
        \draw[edge] (Y) to (X2);
        \draw[edge] (X2) to (R);
    \end{tikzpicture}
\end{minipage} &
\end{tabular}
\caption{Two examples of predictions, each where the prediction rule $R_j$ depends only on the feature $X_j$.}
\label{fig:job:sep1}
\end{figure}

A prediction $R_1$ will fail to satisfy Separation because conditioning on whether a user is a programmer, $Y$, does not d-separate whether he or she has visited \texttt{pinterest.com}, $X_1$, and his or her gender, $A$. Clearly, this will be the case regardless of what information we extract from $X_1$. However, any prediction which is a descendent only of whether the user has visited \texttt{stackexchange.com}, $X_2$, will in fact satisfy Separation, because conditioning on $Y$ blocks the path between $X_2$ and $A$.

Thus we can interpret Separation as a criterion which encourages us to use information which depends on $A$ only through the response, $Y$. However, it is not clear that there are many situations in which we observe features which behave as graphical descendants of the response. We are generally interested in using features which temporally precede the observation of $Y$; usually $X$ causes $Y$ and not the other way around. 

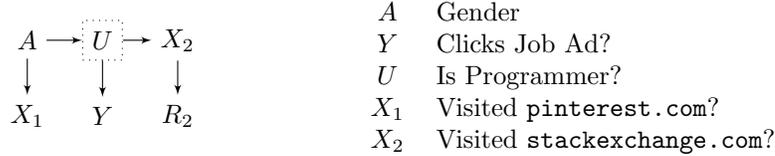
\begin{figure}
\centering
\begin{tabular}{c c}
    \begin{minipage}{.4\textwidth}
    \begin{center}
    \begin{tikzpicture}
        \tikzset{edge/.style = {->,> = latex'}}
        \tikzset{vertex/.style = {shape=rectangle,draw,minimum size=1.5em}}
        \node (A) at  (0,1) {$A$};
        \node (X1) at  (0,0) {$X_1$};
        \node[vertex,dotted] (U) at (1,1) {$U$};
        \node (X2) at (2,1) {$X_2$};
        \node (Y) at (1,0) {$Y$};
        \node (R2) at (2,0) {$R_2$};
        
        \draw[edge] (A) to (U);
        \draw[edge] (A) to (X1);
        \draw[edge] (U) to (X2);
        \draw[edge] (U) to (Y);
        \draw[edge] (X2) to (R2);
    \end{tikzpicture}
    \end{center}
    \end{minipage}
    &
    \begin{minipage}{.4\textwidth}
    \begin{tabular}{c l}
        $A$ & Gender \\
        $Y$ & Clicks Job Ad? \\
        $U$ & Is Programmer? \\
        $X_1$ & Visited \texttt{pinterest.com}? \\
        $X_2$ & Visited \texttt{stackexchange.com}? \\
    \end{tabular}
    \end{minipage}
\end{tabular}
\caption{A perhaps more realistic DAG that underlies the click-predicting task in Scenario 2.}
\label{fig:job:sep2}
\end{figure}

In fact Scenario 2, which was designed to illustrate a possible use of the Separation criterion, is unrealistic. A modification to Scenario 2 which is more realistic can be modelled with the PDAG in Figure~\ref{fig:job:sep2}.

Here, whether or not a user is a programmer is actually an unobserved variable, $U$, and is relevant to us only because it determines the likelihood of the observable event that the user clicks on our advertisement, $Y$. From this more realistic model we can see that $Y$ no longer d-separates $R_2$ from $A$. Thus to the extent that $Y$ is not identical to $U$, Separation will not hold.

\subsubsection{Sufficiency}
\label{scenarios:job:suff}

In contrast to Scenario 1, we cannot construct a non-trivial prediction satisfying Sufficiency. Neither $X_1$ nor $X_2$ are mediators of the effect of $A$ on $Y$. The only way our prediction $R$ could block the direct path between $A$ and $Y$ would be for $R$ to perfectly encode the information in $A$.

\section{Understanding Separation}
\label{understanding}
Among the three oblivious criteria of fairness discussed, we are most skeptical of Separation. As mentioned in Section \ref{measures:sep}, there is a significant literature focused on applying and generalizing the criterion. However, unlike the other criteria, Separation has fundamental limitations, which we now explore.

\begin{figure}
\centering
\begin{minipage}{.4\textwidth}
\begin{center}
\begin{tikzpicture}
    \tikzset{edge/.style = {->,> = latex'}}
    \node (A) at  (2,2) {$A$};
    \node (X1) at  (1,1) {$X_1$};
    \node (X2) at (2,1) {$X_2$};
    \node (X3) at (3,1) {$X_3$};
    \node (Y) at (2,0) {$Y$};
    \draw[edge] (A) to (X1);
    \draw[edge] (A) to (X2);
    \draw[edge] (A) to (X3);
    \draw[edge] (X2) to (X1); 
    \draw[edge, bend right] (X3) to (X1);
    \draw[edge] (X1) to (Y);
    \draw[edge] (X2) to (Y);
\end{tikzpicture}
\end{center}
\end{minipage}
\caption{An example of a DAG where all features are mediators between the sensitive characteristic $A$ and the response $Y$.}
\label{fig:typical}
\end{figure}
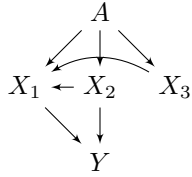

In Section~\ref{scenarios:job:sep}, we found that predictions $R$ which are a function of features that are descendants of the response $Y$ will satisfy Separation, so we will now focus on other cases. In particular, we will focus on an arrangement of the features, sensitive characteristic, and response which we feel is most likely to occur in practice. We assume the DAG is arranged so that the sensitive characteristic $A$ is a root, the response $Y$ is a leaf, and none of the features are descendants of the response. We feel that this DAG is ubiquitous since predictions are often made about the future, so that the features will need to be causal ancestors of the response. An example of such an arrangement is provided in Scenario 1. To be concrete, we will focus on a typical example of a DAG which encodes dependencies between the features, visualized in Figure~\ref{fig:typical}.

We will let $A$ be a root node which may be an ancestor of any of the features, $X_1, \dots, X_p$. Furthermore, features may be ancestors or descendants of each other, but all of the features are ancestors of the response, $Y$.

\begin{figure}
    \begin{minipage}{.4\textwidth}
    \begin{center}
    \begin{tikzpicture}
        \tikzset{edge/.style = {->,> = latex'}}
        \node (A) at  (2,2) {$A$};
        \node (X1) at (1,1) {$X_1$};
        \node (X2) at (2,1) {$X_2$};
        \node (X3) at (3,1) {$X_3$};
        \node (R) at (3,0) {$R$};
        \node (Y) at (2,-1) {$Y$};
        \draw[edge] (A) to (X1);
        \draw[edge] (A) to (X2);
        \draw[edge] (A) to (X3);
        \draw[edge] (X2) to (X1);
        \draw[edge, bend right] (X3) to (X1);
        \draw[edge] (X1) to (R);
        \draw[edge] (X2) to (R);
        \draw[edge] (X3) to (R);
        \draw[edge] (X1) to (Y);
        \draw[edge] (X2) to (Y);
    \end{tikzpicture}
    \end{center}
    \end{minipage}
    \hfill
    \begin{minipage}{.4\textwidth}
    \begin{center}
    \begin{tikzpicture}
        \tikzset{edge/.style = {->,> = latex'}}
        \node (A) at  (2,2) {$A$};
        \node (X1) at  (1,1) {$X_1$};
        \node (X2) at (2,1) {$X_2$};
        \node (X3) at (3,1) {$X_3$};
        \node (th) at (2,0) {$\theta$};
        \node (R) at (3,0) {$R$};
        \node (Y) at (2,-1) {$Y$};
        \draw[edge] (A) to (X1);
        \draw[edge] (A) to (X2);
        \draw[edge] (A) to (X3);
        \draw[edge] (X2) to (X1);
        \draw[edge, bend right] (X3) to (X1);
        \draw[edge] (X1) to (R);
        \draw[edge] (X2) to (R);
        \draw[edge] (X3) to (R);
        \draw[edge] (X1) to (th);
        \draw[edge] (X2) to (th);
        \draw[edge] (th) to (Y);
    \end{tikzpicture}
    \end{center}
    \end{minipage}
    \caption{The leftmost graph is the same example graph, and the rightmost graph shows the signal parameter $\theta$.}
    \label{fig:with.theta}
\end{figure}
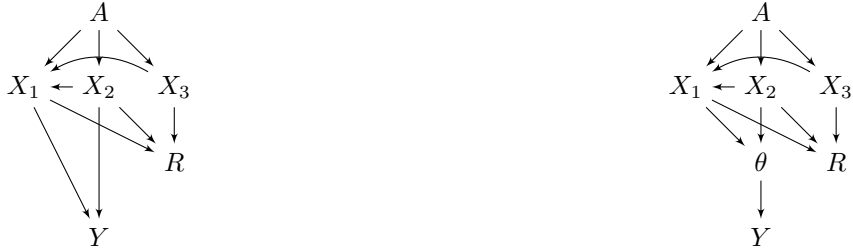

We will now argue that Separation will tend not to hold by examining the structure of the typical DAG, which is visualized with a prediction $R$ in the leftmost graph of Figure~\ref{fig:with.theta}.
There, conditioning on a leaf of the graph, the response $Y$, will not generally block paths from the root, the sensitive characteristic $A$, to the prediction $R$, which depends on the mediators, the features $X$. This is due to a nondegeneracy of the graph: the response $Y$ will be influenced by \emph{exogenous random noise}, say $\epsilon$, in addition to a \emph{signal} through the features $X_1, \dots, X_p$. Due to the interference of the noise $\epsilon$, the information in the response $Y$ is fundamentally different than the information in the features, which leads to the impossibility of blocking, and the failure of Separation.

This argument is easiest to see in the case of the visualized Figure~\ref{fig:with.theta}, but holds more generally. Whenever there is at least one feature which is an ancestor of the response $Y$: the exogenous noise $\epsilon$ will still interfere with the response $Y$, leading to an inability to recover the information in the feature. Specializing the conclusion of this argument to the binary case uncovers a peculiarity in Equalized Odds, despite that Equalized Odds is derived from common measures for summarizing the accuracy of a classifier. 

\subsection{Incompatibility between Separation and Sufficiency}
\label{understanding:incompatability}

Measures of fairness were beginning to be intensely studied and debated when \citet{kleinberg2016inherent} and \citet{chouldechova2017fair} established the surprising result that Calibration and Equalized Odds cannot simultaneously hold in all but degenerate settings. Specifically, it was established:
\begin{theorem}
\label{klein-chould}
    Consider the binary setting, where the response $Y \in \{0,1\}$ and the prediction $R \in \{0,1\}$. When $R \neq Y$, the prediction $R$ can only satisfy both Equalized Odds and Calibration by Group if $P(Y=1 \mid A = a) = P(Y=1 \mid A=a')$ for all $a,a' \in \mathbb{A}$, i.e., the mean response does not differ between levels of the protected characteristic $A$.
\end{theorem}
\noindent The theorem has been generalized beyond the binary case to hold for Separation and Sufficiency. \citet{barocas2017fairness} provide an argument using undirected graphs, which we now reproduce. In \autoref{klein-chould}, we assumed that $R \neq Y$. In the general case, we similarly, but more generally, assume that all events in the joint distribution of $(A, R, Y)$ have positive probability. This assumption makes it so that there's no degeneracy between the random variables. Consider an undirected graphical model of the variables $A$, $R$ and $Y$.  
    
\begin{figure}
    \centering
    \begin{minipage}{.4\textwidth}
    \begin{center}
    \begin{tikzpicture}
        \tikzset{edge/.style = {->,> = latex'}}
        \node (A) at (0,1) {$A$};
        \node (R) at (1,0) {$R$};
        \node (Y) at (0,0) {$Y$};
        \draw[dashed] (A) to (R);
        \draw[dashed] (A) to (Y);
    \end{tikzpicture}
    \end{center}
    \end{minipage}
    \caption{A depiction of the prohibited paths between the random variables $A, R,$ and $Y$ under Separation and Sufficiency.}
    \label{fig:undirected}
\end{figure}
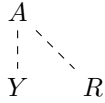

\noindent Separation requires that $R \ind A \mid Y$, so that there can be no edge between $A$ and $R$. Similarly, Sufficiency requires that $Y \ind A \mid R$, so that there can be no edge between $A$ and $Y$. We indicate the impossibility of an edge with a dashed line. Therefore, it must hold that $A \ind Y$, since there can be no path drawn connecting $A$ and $Y$. This establishes the result: under the non-degeneracy assumption, if Separation and Sufficiency both hold, it must be that $A \ind Y$. The condition $A \ind Y$ is a generalization of the condition in \autoref{klein-chould} that the mean response does not differ between levels of the protected characteristic $A$. 

These results cast the disagreement between ProPublica \citep{angwin2016machine} and Northpointe \citep{flores2016false} in a new light. On one hand, the failure of the COMPAS algorithm to achieve equal error rates between groups does not seem to be an objective form of unfairness if it is mathematically impossible for a calibrated classifier to do so. On the other hand, there remains the question of whether disparate impact caused by unbalanced error rates is sufficient cause to dispense with calibration. 

In binary classification settings, it is common to characterize the performance of a classifier using the false positive and false negative rate. These quantities specify the distribution of $R$ given $Y$. However, as we have seen using the example of a simple linear regression model in Section \ref{scenarios:loan:sep}, attempting to enforce parity between quantities conditional on $Y$ can lead to counterproductive procedures. Furthermore, as we have argued, even when Calibration by Group does not hold, in general there is no reason to expect conditioning on $Y$ to block the paths between $R$ and $A$. Thus we believe that the choice between Calibration by Group and Equalized Odds is not a mere subjective trade-off; instead we find Separation to be a fundamentally unhelpful fairness criterion. 

\subsection{Parity by Signal}
\label{understanding:schmeparation}

The original motivation of Equalized Odds \citep{hardt2016equality} was to overcome limitations of Independence. In addition to the stringency of the criterion, the authors argue that a limitation of the criterion is that the response $Y$ itself does not satisfy Independence whenever there is dependence between $Y$ and the sensitive characteristic $A$. This is undesirable, they write, since the response $Y$ is an ``ideal [prediction], which can hardly be considered discriminatory as it represents the actual outcome.''

This line of reasoning seems to obscure a crucial point. When the probability distribution of the response $Y$ depends on the features $X_1, \dots, X_p$ only through a signal parameter $\theta(A,X_1, \dots, X_p)$, which we could take without loss of generality to be the conditional mean $E[Y \mid A,X_1, \dots, X_p]$ when $\theta$ is one-dimensional, the signal $\theta$ will not be an allowable prediction under Separation.
\noindent This follows by the same reasoning as in Section \ref{understanding}: in non-degenerate settings, conditioning on the response $Y$ will not block the dependence between any of its ancestors. This is a significant limitation since the discrepancy between the response $Y$ and the signal $\theta$ is generally unique to each individual and cannot be predicted. Indeed, no prediction rule $R$ can achieve zero prediction error when the response $Y$ has an exogenous noise component. Therefore, in practicality, the perfect prediction $R$ is the signal $\theta$, not the response $Y$. With this in mind, we explicitly define a new measure of fairness.
\begin{definition}
    Represent $Y = F_{\theta(A,X_1,\dots,X_p)}(\epsilon)$, where $\epsilon$ is exogenous noise, that is, $\epsilon \ind A, X_1, \dots, X_p$, and $F$ is a function indexed by a signal parameter $\theta$. Then a prediction $R$ satisfies \emph{Parity by Signal} when $R$ is conditionally independent of the sensitive attribute $A$ given the signal $\theta$, i.e. $R \ind A \mid \theta$.
\end{definition}
\noindent Another way to view this definition of fairness is that the predictions for similar people do not unnecessarily depend on the sensitive characteristic, where similar people are defined to be those whose features contribute---via the signal $\theta(A,X_1, \dots, X_p)$---in the same way to the outcome. This is related to the measure of fairness described by \citet{dwork2012fairness}, wherein Separation by Signal would be considered as utilizing a perfect similarity metric. In Scenario 1, an optimal prediction rule which satisfies Separation was presented, and it was found to be unusual: however, in that same scenario, the true mean does indeed satisfy Parity by Signal.

This definition of fairness is not without its limitations. Evaluating whether a prediction $R$ satisfies Separation by Signal requires the signal $\theta$, which is generally not known in practice. Above we demonstrated that Separation by Signal is a useful device to develop understanding, but a close variant of it can also be made operational. Separation by Signal compares the prediction $R$ to the signal $\theta$, but, instead, $R$ could be compared to another prediction $S$ which we believe to be more accurate than $R$.
\begin{definition}
    A prediction $R$ satisfies \emph{Parity by $S$} if $R$ is conditionally independent of the sensitive attribute $A$ given $S$, i.e. $R \ind A \mid S$.
\end{definition}
\noindent Notice that a prediction $R$ satisfies Separation by Signal if and only if $R$ satisfies Separation by $\theta$. 

We can interpret a variety of fairness-testing procedures as a form of testing for Parity by $S$. For example, in the context of testing whether various police precincts exhibit racial bias in contraband searches, \citet{simoiu2017problem} develop a threshold test which is a test for a kind of Parity by Signal. They are in the binary setting and consider the prediction $R(X_1, \dots, X_n) = \mathbb{I} \left[ p_A(X_1, \dots, X_p) > t_A \right]$ to be that an individual is carrying contraband when the probability $p_A(X_1, \dots, X_p)$ of the individual carrying contraband is larger than a threshold $t_A$ and to be that an individual is not carrying contraband otherwise. They develop Bayesian tests for whether the threshold $t_A$ depends on the race $A$, since they argue that a fundamental form of unfairness occurs when minorities are ruled against more stringent thresholds. Due to the prediction $R$ depending only on the probability $p_A$ and the threshold $t_A$, this is a test for whether $R$ satisfies Parity by $p_A$.

In the above example, the threshold test sought to determine whether there was a specific form of bias in police officers' decisions to search for contraband. This is an example of testing subjective human predictions, with some modeling assumptions. However, a Separation by $S$ criterion can also be desirable to hold for a prediction $R$ even when both $R$ and $S$ are generated by machine algorithms.
Consider cases in which we believe that a model is unfair due to misspecification; perhaps this model is missing necessary features or fails to model interactions between the sensitive characteristic and other features in a way which leads the predictions generated by the algorithm to disparately impact certain groups. (See in particular Scenario 1: The Red Car in \citet{kusner2017counterfactual}.) Specifically, suppose that the prediction $R(X) = \beta^T X$ and the prediction $S(X) = \beta_A^T X$, where each coefficient $\beta_A$ differs based on the sensitive characteristic $A$. In this case, a likelihood ratio test \citep{agresti2015foundations} between these models would be a test for whether $R$ satisfies Separation by $S$.

\section{Causal Considerations}
\label{causal}

In various discussions of Independence and Separation-like criteria, authors propose generalizations which enforce parity only after conditioning on certain features \citep{barocas2017fairness,hardt2016equality,donini2018empirical}. Consider the following generalizations of Independence. Suppose $X$ is some subvector of the features $X_1, \dots, X_p$.
\begin{definition}
\label{cond_ind}
    A prediction $R$ satisfies Conditional Independence with respect to $X$ if $R \ind A \mid X$.
\end{definition}
Notice that Conditional Independence with respect to $(X_1, \dots, X_p)$ always holds when the prediction $R$ is a deterministic function of the features $X_1, \dots, X_p$. This reinforces that the purpose of Conditional Independence is to study the influence of a subvector $X$ of the features. Notice also the connection to Parity by Signal and Parity by a prediction $S$, discussed in Section~\ref{understanding:schmeparation}, which involve evaluating independence conditional on a specific function of the features.

Conditional Independence criteria themselves convey little information about the underlying desires of the practitioner. However, causal reasoning can provide principled methods for developing fairness criteria which may ultimately be expressed as Conditional Independence criteria. Here, we discuss two scenarios in which Conditional Independence criteria can be derived using causal reasoning. For a variety of perspectives on the role of Pearl's causality theory in fairness, see \citet{kusner2017counterfactual,kilbertus2017avoiding,nabi2018fair} and \citet{chiappa2018path}.

\subsection{Scenario 3: College Admissions}
\label{causal:college}

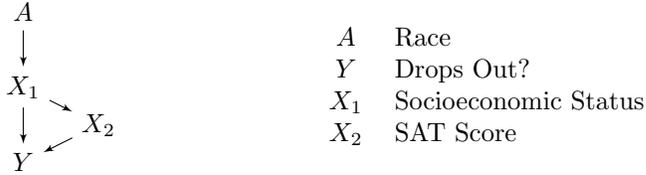
\begin{figure}
\begin{tabular}{c c}
    \begin{minipage}{.4\textwidth}
    \begin{center}
    \begin{tikzpicture}
        \tikzset{edge/.style = {->,> = latex'}}
        \node (A) at  (0,2) {$A$};
        \node (X1) at  (0,1) {$X_1$};
        \node (X2) at (1,.5) {$X_2$};
        \node (Y) at (0,0) {$Y$};
        \draw[edge] (A) to (X1);
        \draw[edge] (X1) to (X2);
        \draw[edge] (X1) to (Y);
        \draw[edge] (X2) to (Y);
    \end{tikzpicture}
    \end{center}
    \end{minipage}
    &
    \begin{minipage}{.4\textwidth}
    \begin{tabular}{c l}
        $A$ & Race \\
        $Y$ & Drops Out? \\
        $X_1$ & Socioeconomic Status \\
        $X_2$ & SAT Score \\
    \end{tabular}
    \end{minipage}
\end{tabular}
\caption{The random variables in Scenario 3: various features and their relationship to race $A$ and drop out status $Y$.}
\label{fig:college}
\end{figure}

For the purpose of college admissions, we wish to predict whether a student will drop out before completing his or her degree. Suppose we model the situation using the PDAG in Figure~\ref{fig:college}.

Suppose that the admissions committee wishes to use all of the relevant information about student performance contained in the student's SAT score $X_2$, but otherwise wishes to ignore the student's race-laden socioeconomic status $X_1$, despite the fact that socioeconomic factors do have a direct effect on a student's probability of dropping out. In the language of \citet{kilbertus2017avoiding}, this means that $X_2$ is a \emph{resolving} variable. The Independence criterion would not allow us to use all of the information in $X_2$, since we would have to extract the component of $X_2$ which is independent of the student's race $A$. Separation or Parity by Signal, on the other hand, would not allow us to fully make use of $X_2$ while entirely excluding information in $X_1$.

In fact we actually desire for $A$ to have no \emph{direct effect} on $R$ that is not mediated by $X_2$. We can express this criterion using the formula for Controlled Direct Effect (CDE) as:
\begin{equation}
\label{CDE}
    \mathcal{P}(R \mid \text{do}(A = a),\text{do}(X_2 = x)) = \mathcal{P}(R \mid \text{do}(A = a'),\text{do}(X_2 = x))
\end{equation}
for all $a,a' \in \mathbb{A}$ and all $x$ in the range of $X_2$. In this case, this do-expression is identifiable and simplifies to the expression $R \ind A \mid X_2$. However, in cases when a resolving variable is itself confounded with other variables, the do-expressions in (\ref{CDE}) may be unidentifiable or require some do-calculus to resolve into observational expressions.  

\subsection{Scenario 4: Insurance Prices}
\label{causal:insurance}
In the scenarios we have discussed so far, we model the sensitive characteristic $A$ as an exogenous variable. Thus $A$ has always been a root in our PDAG models, and the total causal effects of $A$ on endogenous entities such as $R$ and $Y$ coincide with the observed effects. However, when $A$ is an endogenous variable, there may be backdoor paths from $A$ to our predictions or response. Consider the following scenario, which makes clear the need for causal reasoning in fairness.

\begin{figure}
\begin{tabular}{c c}
\begin{minipage}{.4\textwidth}
\begin{center}
    \begin{tikzpicture}
        \tikzset{edge/.style = {->,> = latex'}}
        \tikzset{vertex/.style = {shape=rectangle,draw,minimum size=1.5em}}
        \node[vertex,dotted] (U) at (1,1) {$U$};
        \node (A) at (0,0) {$A$}; 
        \node (Y) at (1,0) {$Y$};
        \node (X) at (2,0) {$X$};
        \draw[edge] (U) to (A);
        \draw[edge] (U) to (Y);
        \draw[edge] (U) to (X);
    \end{tikzpicture}
\end{center}
\end{minipage}
&
\begin{minipage}{.4\textwidth}
    \begin{tabular}{c l}
        $A$ & Religion \\
        $Y$ & Car Accident? \\
        $U$ & Personality \\
        $X$ & Past Traffic Tickets 
    \end{tabular}
\end{minipage}
\end{tabular}
\caption{The random variables in Scenario 4: an unobserved variable, and various features and their relationship to religion $A$ and car accident status $Y$.}
\label{fig:insurance}
\end{figure}
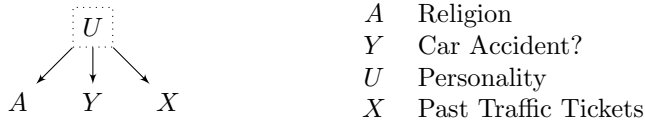

We wish to predict whether an individual is likely to have a car accident for the purpose of determining her insurance premium. Suppose we model the relevant variables using the PDAG in Figure~\ref{fig:insurance}.

We may deem it unfair to use an individual's religion, $A$, as a factor in our prediction, $R$. However, there is a backdoor path between $A$ and $Y$, and thus $A$ and $Y$ are marginally dependent. For the same reason, any prediction $R$ which is a non-trivial descendent of the individual's driving record $X$ will fail to satisfy Independence. Of course, as in all of the ubiquitous cases when the features are ancestors of the response, conditioning on $Y$ will not black the path between $R$ and $A$ either, thus our prediction will also fail to satisfy Separation.

In this case, we need not consider $X$ to be a resolving variable. Whatever dependence results between $A$ and $R$ is spurious. We are interested in ensuring that $A$ has no Total Effect (TE) on $R$, which we can express as:
\begin{equation}
\label{TE}
    \mathcal{P}(R \mid \text{do}(A = a)) = \mathcal{P}(R \mid \text{do}(A = a')).
\end{equation}
 Note that this is a special case of the Counterfactual Fairness criterion in \citet{kusner2017counterfactual}, although these authors do not explicitly consider cases in which $A$ is endogenous. The consequence of this criterion is that we can freely construct predictions $R$ which are based on traffic tickets, $X$, or other inferred aspects of personality, $U$.

\section{Conclusion}
\label{conclusion}

In Scenarios 1 through 4, PDAGs have proven to be fertile ground for developing intuition about the three basic oblivious criteria for fairness: Independence, Separation, and Sufficiency. In general, constructing a PDAG model relating the sensitive characteristic, features, and response is a clarifying exercise, because it allows us to more directly connect our senses about what is intuitively fair to the implications of the decisions we make in specifying an algorithm. In contrast to the project of constructing statistically optimal estimators, a fundamental concern in constructing fair estimators is blocking the use of information which is subjectively unacceptable. Here, PDAGs and d-separation are natural tools.

In contrast, the oblivious fairness criteria alone are limited, because their behavior is opaque and sensitive to the particularities of the scenario. Enforcing Independence between the sensitive characteristic and the prediction was seen to have wildly different implications in scenarios when the response was dependent on race and when it was not. In the former case, Independence prohibited 
discrimination which could not be statistically justified, and in the latter case, Independence was an intervention in favor of adversely affected groups.

Sufficiency can be achieved by blocking all paths through which information can flow between the sensitive characteristic and the response. Generally this is only possible by accurately recovering the signal. That is, it was shown that Sufficiency can be achieved by appropriately choosing the features through which information flows from race to the response and appropriately choosing a prediction that blocks that flow of information. 

Separation naturally allows for the use of features which are descendants of the response, but exhibits strange behavior whenever there are features which are non-descendants of the response, even when those features are independent of the sensitive characteristic. For non-degenerate and faithful PDAG models, Separation will not hold since the response is comprised of not only the signal, but also the noise, which obscures the information about the signal in the response. In Scenario 1, a violation of faithfulness was induced to produce an optimal prediction rule that turned out to be highly inappropriate, for some groups even requiring the addition of further random noise.

For the most part, these measures have been found wanting. 
We join the recent consensus that the assessment of fairness in algorithms should not start and end with the use of a singular criterion. The constraints we wish to impose on predictions should be sensitive to each scenario, and PDAG models can help to explore them.

More generally, we notice that there is little consensus on the underlying philosophical principles that should provide the foundation for the quantification of fairness. While much work in fairness seems to be framed around preventing unfairness like that allegedly exhibited by the COMPAS algorithm, there is no consensus that the COMPAS algorithm was ever unfair. 
We worry that as constructs from fairness are taken out of context and treated as black boxes for mathematical study and elaboration, the implicit underlying notions of fairness will be obscured.
We hope that fairness research can be grounded in clear, practical examples of the ways algorithms can be unfair.

\section*{Acknowledgements}
\label{acknowledgements}

Thanks to David Kent for helpful conversations.

\newpage
\bibliography{bib}
\bibliographystyle{plainnat}
\end{document}